\documentclass{article}

\usepackage{PRIMEarxiv}
\usepackage[title]{appendix}
\usepackage[utf8]{inputenc} 
\usepackage[T1]{fontenc}    
\usepackage{hyperref}       
\usepackage{url}            
\usepackage{amsfonts}       
\usepackage{nicefrac}       
\usepackage{microtype}      
\usepackage{lipsum}
\usepackage{physics}
\usepackage{fancyhdr}       
\usepackage{graphicx}       
\graphicspath{{media/}}     
\usepackage{amsmath}
\usepackage{subcaption}

\title{Swirl flow in microchannels: patterned slip walls enhance heat transport}

\author{
  L. G. Chej, M. F. Carusela, A. G. Monastra \\
  Laboratorio de Modelado y Simulación Computacional,  Instituto de Ciencias \\
  Universidad Nacional de General Sarmiento \\
  Los Polvorines, Buenos Aires, CP1613, Argentina\\
  Consejo Nacional de Investigaciones Científicas y Técnicas\\
  Ciudad Autónoma de Buenos Aires, C1425FQB, Argentina\\
  Corresponding author:
  \texttt{lchej@campus.ungs.edu.ar} \\
   \And
  J. Harting \\
  Helmholtz-Institut Erlangen-Nürnberg für Erneuerbare Energien (IET-2)\\
  Forschungszentrum Jülich\\
  Cauerstraße 1, 91058 Erlangen, Germany\\
  Department Chemie—und Bioingenieurwesen und Department Physik\\
  Friedrich-Alexander-Universität Erlangen-Nürnberg\\
  Cauerstraße 1, 91058 Erlangen, Germany
  \And
  P. Malgaretti\\
  Helmholtz-Institut Erlangen-Nürnberg für Erneuerbare Energien (IET-2)\\
  Forschungszentrum Jülich\\
  Cauerstraße 1, 91058 Erlangen, Germany
}

\begin{document}
\maketitle

\begin{abstract}
Microchannel heat sinks (MCHS) are widely used for thermal management in high-power electronics due to their ability to dissipate large heat fluxes with minimal coolant consumption. While numerous strategies—such as geometric modifications, surface disruptions, and enhanced coolant formulations—have been explored to improve heat transfer, many of these approaches increase hydraulic resistance and pumping power requirements. Recent studies have shown that slip/no-slip wall patterns can enhance flow rates and convective heat removal without additional energy input, and that patterned microstructures can induce secondary swirling motions known to promote mixing and heat transfer. Motivated by these findings, we investigate a slip/no-slip pattern specifically designed to generate swirl flow inside a straight microchannel. Building upon prior work on passive chaotic advection and boundary-condition engineering, we assess the hydrodynamic and thermal performance of this patterned configuration under conditions relevant to laminar microchannel cooling. Our results demonstrate that appropriately arranged slip/no-slip regions can induce swirl without geometric perturbations or increased pumping power, ultimately improving heat transfer efficiency at fixed volumetric flow rate. This study highlights the potential of boundary-condition patterning as a simple, energy-neutral strategy for enhancing the performance of microfluidic heat-transfer devices.
\end{abstract}


\clearpage

\section{Introduction}
Since the seminal work of Tuckerman and Pease \cite{tuckermanHighperformanceHeatSinking1981}, microchannel heat sinks (MCHS) have been established as an efficient solution for dissipating large heat fluxes—up to \(790~\text{W}/\text{cm}^2\)—using minimal coolant. In recent years, the trend toward smaller transistor sizes and higher chip densities has increased the thermal load in microprocessors \cite{bahiraeiElectronicsCoolingNanofluids2018}, reinforcing the relevance of microchannel cooling technologies. MCHS are typically fabricated from highly thermal conductive materials such as silicon, aluminium, or copper, and most often employ liquids (e.g., water or ethylene glycol) as coolants.

The main objective of these devices is to maximize heat removal through forced convection at the outlet. However, a fraction of the input heat—typically around \(10\%\) \cite{khoshvaght-aliabadiExperimentalStudyCooling2016a,leeInvestigationHeatTransfer2005}—is inevitably lost through mechanisms such as convective exchange with ambient air, thermal radiation, and conduction through the device’s top wall. For optical accessibility during diagnostics and temperature measurements \cite{jungHeatTransferFlow2019,chaudhuriDiscreteElectricField2017,timungElectricFieldMediated2017}, this wall is often made from insulating transparent materials like glass, PDMS, or PMMA \cite{linPDMSMicrofabricationDesign2021}.

A key parameter governing hydrodynamic behavior is the aspect ratio \(L/w\),
which typically ranges between 62 and 210 \cite{quExperimentalNumericalStudy2002b, harms1999}. Heat removal efficiency depends on the volumetric flow rate \(\dot{V}\), with most devices operating under laminar conditions characterized by low Reynolds numbers \(Re = \bar{u}_z D_h / \nu\). Although some studies explore regimes up to the laminar–turbulent transition at \(Re \sim 2500\) \cite{leeInvestigationHeatTransfer2005}, most operate within \(Re = 30\text{–}100\) \cite{kimExperimentalStudyFully2016a,sohelExperimentalInvestigationHeat2014}.

\begin{figure}[hbtp]
    \begin{center}
    \includegraphics[width=0.7\linewidth]{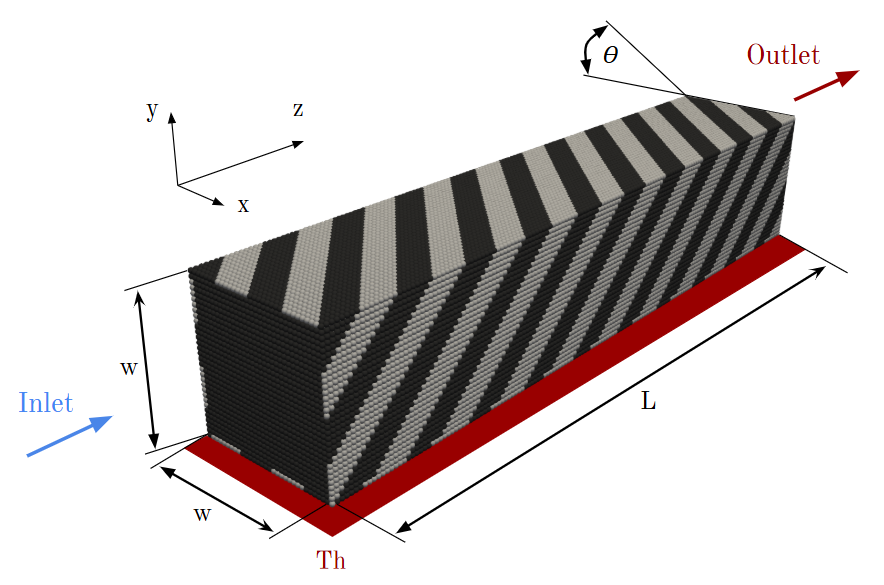}
    \caption{Square duct representation with a slip/no-slip pattern in the channel walls.}
    \label{fig:slip_pattern}    
    \end{center}
\end{figure}

Performance improvements in MCHS have been pursued through variations in channel geometry and coolant composition, including nanofluids \cite{sohelmurshedCriticalReviewTraditional2017, kaltehExperimentalNumericalInvestigation2012}. Numerous cross-sectional shapes—trapezoidal, circular, triangular, among others—and structured perturbations such as sinusoidal or ribbed channels have been investigated \cite{mohammedadhamThermalHydrodynamicAnalysis2013}. More recently, swirling flows have been shown to enhance heat transfer \cite{fanHeatTransferPerformance2024,fengNumericalInvestigationLaminar2017a}, often generated using ribs, obstacles, or other geometric disruptions \cite{sheikholeslamiReviewHeatTransfer2015a,kumarReviewHeatTransfer2018}. While effective, these modifications usually increase the required pumping power.

Interestingly, it has been shown that patterned grooves can induce chaotic advection, generating vortical structures\cite{stroockChaoticMixerMicrochannels2002,sarkarQuantificationPerformanceChaotic2012a,gluzdovSlipLengthAnalysis2024a}.
Moreover, slip/no-slip patterns have been used to induce swirl flows: 
Ouroo-Koura et al. \cite{ouro-kouraBoundaryConditionInduced2022} examined their role in passive mixing, while Rahbarshahlan et al. \cite{rahbarshahlanNumericalSimulationFluid2020} and Babaei et al. \cite{babaeiNumericalStudyHeat2023}, building on Antman et al. \cite{antmanMicroflowsNanoflows2005}, analyzed their thermal implications. These works show that such patterns can enhance the flow rate at constant pumping power and improve heat removal through forced convection.

In this paper, we extend these efforts by investigating a slip/no-slip pattern specifically designed to induce swirl flow within the microchannel. Our aim is to enhance heat transfer without increasing the volumetric flow rate by increasing the applied pressure.

\section{Physical Model}\label{sec:physical model}

The physical model consist on a square duct of side $w=5\times10^{-4}$m and length $L=50\times w$. The working fluid is a water ethylene-glycol mixture (75$\%$) \cite{bohneThermalConductivityDensity1984}. Its thermophysical properties at 25 °C are: 
density $\rho=1088$ kg/m$^3$, kinematic viscosity $\nu=6.296\times 10^{-6}$m$^2$/s, specific heat capacity $c_p$ = 2768 J/(kg K), thermal conductivity $\kappa = 0.307$ W/(mK) and thermal diffusivity $\alpha = 1.018\times 10^{-7}$ m$^2$/s. The flow is driven by a pressure drop $\Delta$P ranging from 0.3 kPa to 10 kPa. These parameters ensure a laminar flow with a Reynolds number below 50, and a thermal transport regime dominated by advection over diffusion, with a Peclet number $Pe = w \bar{u}_z / \alpha$ over 300.

The system is heated at the bottom wall with a hot temperature $T_h = 35^{\circ}$C. The inlet coolant fluid temperature is always $T_c = 15^{\circ}$C, being also the initial temperature of the fluid along the channel. The others three walls of the square duct have an adiabatic boundary condition. Given these conditions, it is expected that in the stationary regime, the temperature of the fluid ranges between $15^{\circ}$C $< T < 35^{\circ}$C. 

The square duct walls are decorated by stripes, tilted by an angle $\theta$ with respect to the flow direction, with different wetting properties. This leads   to a patterned slip/no-slip condition on the walls captured by the  Maxwell-Navier slip length $b$ that can be interpreted as a fictitious distance underneath the solid surface where the no-slip boundary condition would be satisfied \cite{laugaMicrofluidicsNoSlipBoundary2007}. To achieve a full slip boundary condition a slip length $b \rightarrow \infty$ is considered. 
The number of slip and no-slip stripes in a duct along the channel is $n$, while $L/n$ is the width of each stripe.
From here on, for different slip pattern combinations, half of the total duct area has a no-slip boundary condition and the other half has a full slip boundary condition (grey/black stripes in Fig. \ref{fig:slip_pattern} respectively). 

To study the dynamic and thermal transfer behavior of the model, it is necessary to consider the set of conservation equations for the fluid:

Conservation of mass:
\begin{equation} \label{eq:mass}
    \pdv{\rho}{t}+\mathbf{\nabla} \cdot (\rho \mathbf{u})= 0
\end{equation}
where $\rho$ is the mass density of the fluid and $\mathbf{u} = u\hat{i} + v\hat{j} + w\hat{k}$ is the fluid velocity.

Conservation of momentum:
\begin{equation} \label{eq:momentum}
    \rho \frac{D\mathbf{u}}{Dt}=\mathbf{\nabla} \cdot \overline{\overline{\sigma}} +\rho \mathbf{d}\,,
\end{equation}
$\mathbf{d}$ represents the external forces per unit mass and $\overline{\overline{\sigma}}$ the deviatoric stress tensor.

Conservation of energy:
\begin{equation} \label{eq:energy}
    \rho \frac{De}{Dt}=-p \mathbf{\nabla}\cdot\mathbf{u}-\mathbf{\nabla}\cdot\mathbf{q}+\Phi\,,
\end{equation}
where $e$ corresponds to the internal energy of the fluid per unit mass and $\Phi$  is the dissipated kinetic energy.
The first term on the right hand side of Eq.~\eqref{eq:energy} represents the work done by the external pressure $p$ on the fluid. The second term $\mathbf{q}$ is the heat flux
\begin{equation}
    \mathbf{q} = - \kappa \mathbf{\nabla} T + \mathbf{q}_{\text{rad}} ,
\end{equation}
with its first term due to heat conduction (Fourier's law) and the second one due to electromagnetic radiation. $\kappa$ is the thermal conductivity of the coolant. 

To simplify the general equations, we consider a Newtonian and incompressible fluid, with constant thermophysical properties ($\rho$, $\nu$, $c_p$ and $\kappa$). We neglect the viscous dissipation $\Phi$ and $\textbf{q}_{rad}$, and we consider a linear dependence of the internal energy $e$ with the temperature $T$. Therefore, Eqs. \ref{eq:mass}, \ref{eq:momentum} and \ref{eq:energy} become

\begin{equation}\label{eq:C}
    \mathbf{\nabla}\cdot \mathbf{u} = 0
\end{equation}
\begin{equation}\label{eq:NS}
    \rho \frac{D\mathbf{u}}{Dt} = - \nabla p + \mu \nabla^2 \mathbf{u}
\end{equation}
\begin{equation}\label{eq:AD}
    \frac{\partial T}{\partial t} + \mathbf{\nabla}\cdot(T\mathbf{u}) = \alpha \nabla ^2 T ,
\end{equation}
with $\mu$ being the dynamic viscosity. The grade of swirl is measured through the vorticity number, defined by

\begin{equation}
    \Omega_z = \frac{dv}{dx}-\frac{du}{dy}     \label{eq:vorticity}.
\end{equation}

\section{Computational Method}\label{sec:computational method}
We use the Lattice Boltzmman Method (LBM)\cite{Schiller2017,krugerLatticeBoltzmannMethod2017b}
, to solve the equations \ref{eq:C}, \ref{eq:NS} and \ref{eq:AD} on the computational domain. The method solves the Boltzmman Equation  on a discrete lattice:

\begin{equation}
f_i(\mathbf{x}+\mathbf{c}_i \Delta t,t+\Delta t)-f_i(\mathbf{x},t)=
    \hat{\Omega}(f_i(\mathbf{x},t))\,,
\end{equation}
here $\Delta x$ and $\Delta t$ are the units in which space and time are discretized.
The single particle distribution function $f_i$ describes the probability to find a particle with velocity $\mathbf{c}_i$ reaching the adjacent position $\mathbf{x}+\mathbf{c}_i\Delta t$, one time step $t + \Delta t$ later. 
The collision operator $\hat{\Omega}(f_i)$ is the source term proposed by Bhatnagar, Gross, and Krook (BGK) \cite{bhatnagarModelCollisionProcesses1954}: 
\begin{equation}
    \hat{\Omega}(f_i(\mathbf{x},t)) = - \frac{1}{\tau}\left(f_i(\mathbf{x},t)-f_i^{(eq)}(\mathbf{x},t)\right) \ ,
\end{equation}
which describes how the population 
$f_i$ relaxes towards a local equilibrium over the characteristic relaxation time $\tau$, while conserving mass and momentum where the equilibrium distribution is defined as

\begin{equation}
    f_i^{\mathrm{eq}} = w_i \rho\left[1+\frac{\boldsymbol{u} \cdot \boldsymbol{c}_i}{c_s^2}+\frac{\left(\boldsymbol{u} \cdot \boldsymbol{c}_i\right)^2}{2 c_s^4}-\frac{\boldsymbol{u} \cdot \boldsymbol{u}}{2 c_s^2}\right]\,,
    \label{eq:eq_distr}
\end{equation}
with

\begin{align}
    \rho(\mathbf{x},t) &= \sum_{i=1}^{19}f_i(\mathbf{x},t)\,,\\
    \mathbf{u}(\mathbf{x},t) &= \frac{1}{\rho(\mathbf{x},t)}
    \sum_{i=1}^{19}f_i(\mathbf{x},t)\mathbf{c}_i\,.
\end{align}

We adapted the algorithm proposed in the previous work of Peter~\cite{peterNumericalSimulationsSelfdiffusiophoretic2020} and  Antunes~\cite{antunesPumpingMixingActive2022a}, to  solve the thermal advection-diffusion equation~ (Eq. \ref{eq:AD}),
allowing us to study the heat transfer in the computational fluid domain. The validation of this adapted algorithm in a simple 2D case is done comparing with an analytical solution that is detailed in Appendix~\ref{sec:Appendix II}.

We choose the number of LB nodes for our square duct $w^{\star}=32$, and from now on all the star denoted quantities are in lattice units (lu). The corresponding spatial resolution results in $C_x = w/w^{\star}$. This resolution is chosen after a mesh convergence study (see Appendix~\ref{sec:Appendix II} ), seeking for an optimal compromise between computational cost and accuracy of the results.

The liquid number density is taken as $\rho^{\star}=1$ and the relaxation time (which controls the kinematic viscosity) $\tau^{\star}=0.8$. The lattice speed of sound is $c_s^{\star}=\sqrt{1/3}$. The  temperature in lattice units ranges between $0.1<T^{\star}<0.3$, with a temperature difference of $\Delta T^{\star}=0.2$ within the system.\\
With this variables defined, we can obtain the conversion factor for the time through the kinematic lattice viscosity $\nu^{\star}$ equation (Eq. 7.14 from Ref.~\cite{krugerLatticeBoltzmannMethod2017b})

\begin{equation}\label{eq:time_conv_factor}
    C_t = c_s^{\star 2}\left( \tau^{\star} - \frac{1}{2} \right) \frac{C_x^2}{\nu} .
\end{equation}
The pressure drop is mimicked by applying an external force density which accelerates the fluid along the $z$-direction with $a\in[1,3.5]\cdot 10^{-4}$ in lattice units. 

The thermal diffusivity used in Eq.~\eqref{eq:AD} can be expressed in lattice units as $\alpha^{\star}=\alpha/C_{\alpha}$, with $C_{\alpha}= C_x^2/C_t$.
In Refs \cite{khalidahmed2009,hechtImplementationOnsiteVelocity2010} the slip length $b^{\star}$ in lattice units is related to a slip parameter $\zeta^{\star}$ and relaxation time $\tau$ as

\begin{equation}\label{eq:slip_length}
    b^{\star} =  \frac{ \zeta^{\star}}{3(1-\zeta^{\star})} \tau^{\star}\,.
\end{equation}

To obtain a no-slip boundary condition, in our simulations we set $\zeta^{\star} = 0$ to obtain a slip length tending to zero. Meanwhile, $\zeta^{\star} \rightarrow 1$ is recovers the full slip boundary condition.

\section{Results}\label{sec:results}

We simulate a microchannel subjected to different pressure gradients with the following boundary conditions for the velocity at the walls: (i) homogeneous no slip, (ii) homogeneous full slip, and (iii)  heterogeneous slip pattern. The slip pattern variations are: 25, 50, 100, and 200 stripes at each microchannel wall, and $25^{\circ}$, $45^{\circ}$ and $65^{\circ}$ stripe angles.
For all the simulations, we checked that both the liquid density $\rho^{\star}$ and the volumetric flow rate $\dot{V}^\star$ fluctuates within a range of less than $1\%$ and $2\%$ of their respective mean values.

The temperature of the bottom wall is kept constant at $T=T_h$, generating an inlet heat flow $Q$. Considering the adiabatic nature of the external walls the same $Q$ is removed at the outlet by forced fluid convection, computed as

\begin{equation}
    Q = \int_A c_p \rho u_z (T_{\text{out}} - T_{\text{in}})dA ,
\end{equation}
where $c_p$, $\rho$, and $u_z$ represent the specific heat capacity, density, and axial velocity of the working fluid, respectively. $T_{\text{out}}$ and $T_{\text{in}} =T_c$ are the temperatures at the outlet and inlet, respectively. 

\begin{figure}[h!]
    \centering
    \includegraphics[width=0.45\linewidth]{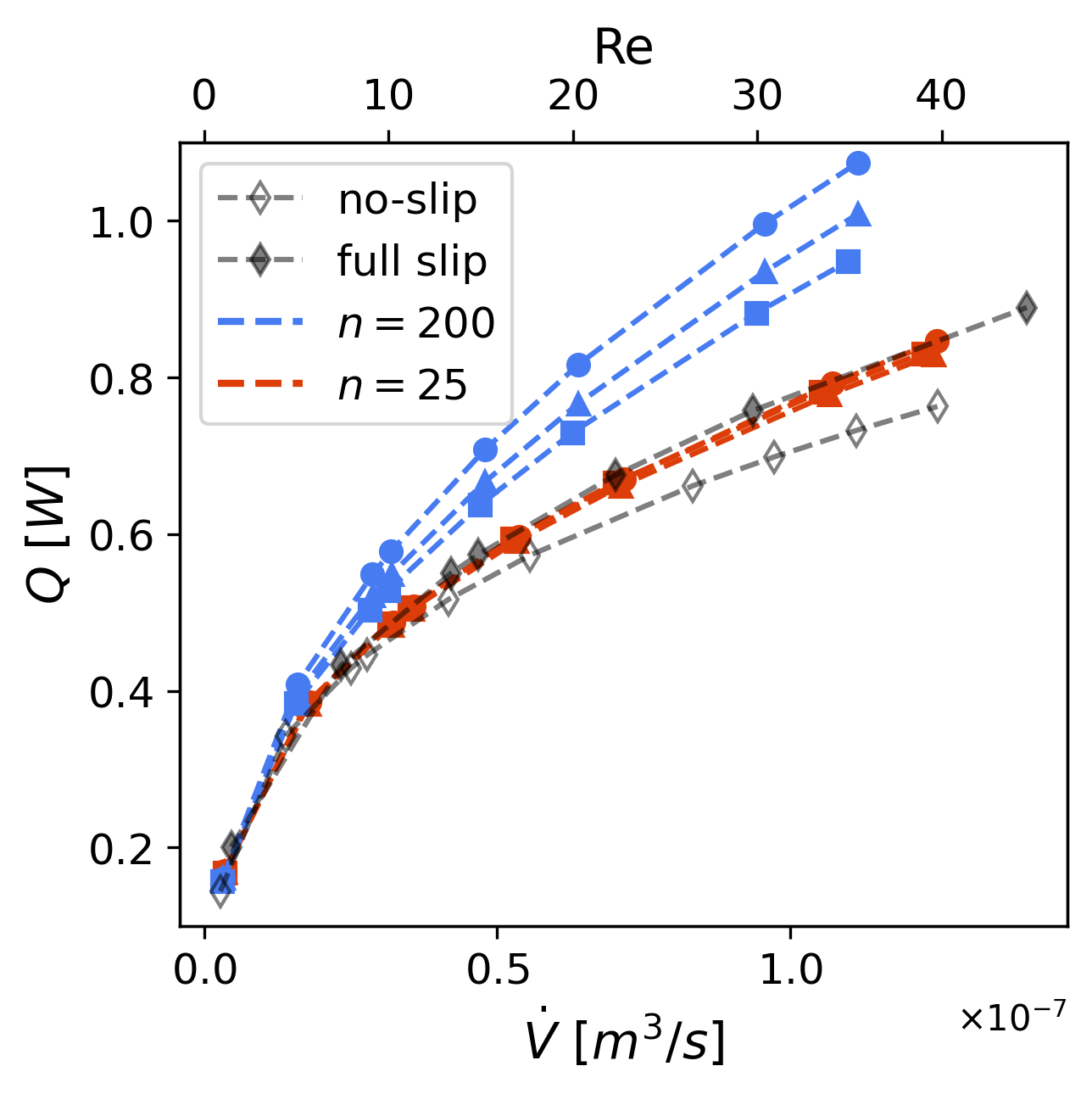}
    \caption{Heat flow evacuated at different flow rate regimes varying the stripe angle $\theta$ at the channel walls (triangles for $\theta=25^{\circ}$, circles for $\theta=45^{\circ}$, and squares for $\theta=65^{\circ}$). Thinner stripes case (largest number of stripes, $n=200$) are the blue ones, and the thicker stripes case (fewer number of stripes, $n=25$) are the red ones. For reference, we plot the homogeneous no-slip case (open diamonds) and the homogeneous full slip case (filled diamonds). The dashed lines between the points serve as a guide for the eye.}
    \label{fig:angles}
\end{figure}

Firstly, we focus on the dependence of the extracted heat $Q$ at the outlet as a function of the volumetric fluid flow, $\dot{V}$ for different covering patterns, as it is shown in Fig. \ref{fig:angles}. It is observed, as expected, that the heat flow $Q$ increases with $\dot{V}$ for all slip patterns as well as for the full slip and no slip cases.  
In Fig. \ref{fig:angles}, we present six different pattern configurations corresponding to $n = 25$ and $n=200$ as well as three stripe angles $\theta = 25^{\circ}, 45^{\circ}, 65^{\circ}$.
As shown in Fig.~\ref{fig:angles}, for the case $n=25$, the three different stripe angles show very similar values of the heat transfer $Q$.
In contrast, for $n=200$, the curves for the three different angles clearly separate, with the case $\theta = 45^{\circ}$ being the one leading to the maximum value of $Q$. 
Such a behavior persist also for interemediate stripes number, $n=50,100$ as shown in Appendix (\ref{sec:Appendix III})

\begin{figure}[htbp]
  \centering
  \includegraphics[width=0.5\linewidth]{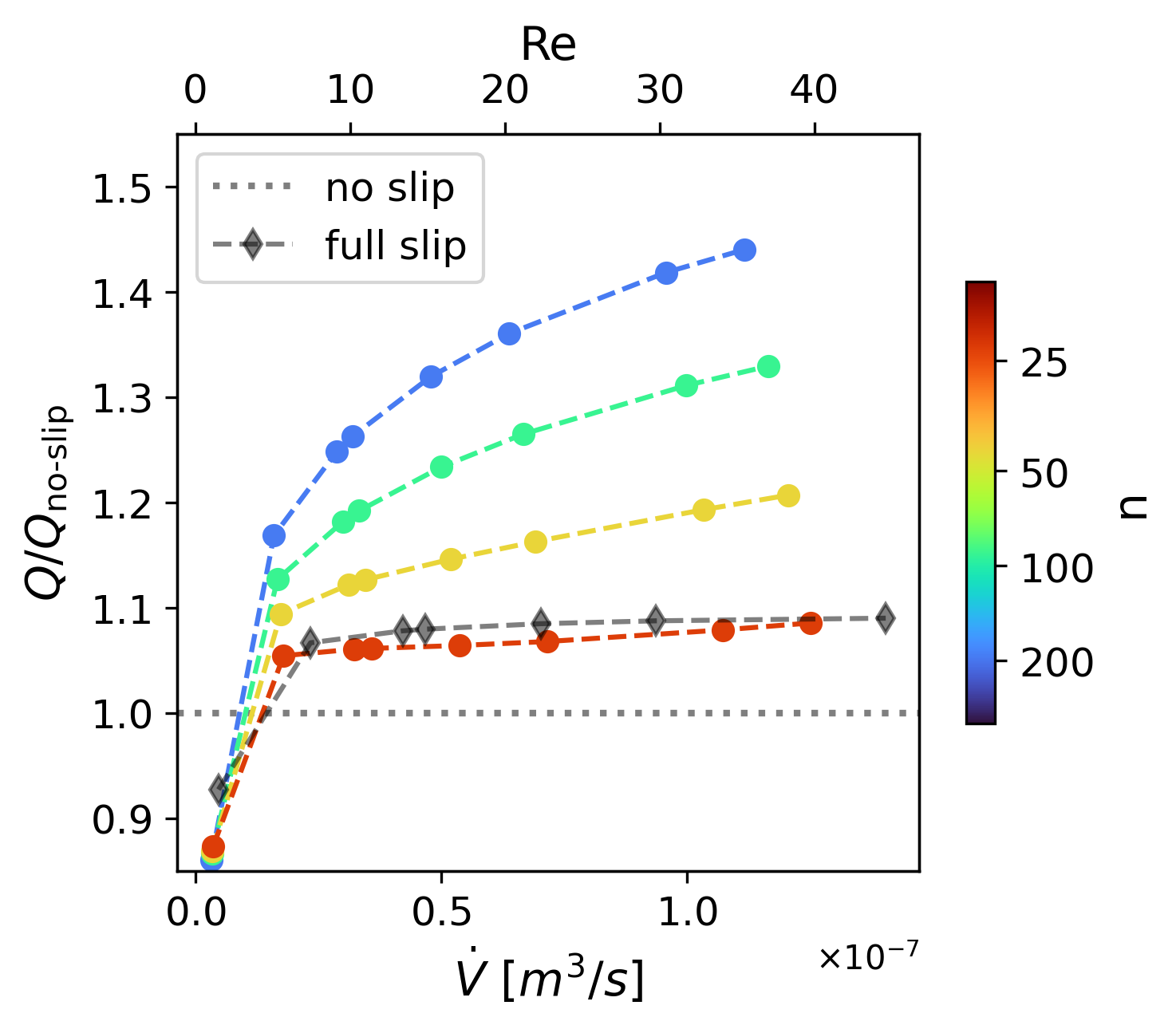}
  \caption{Heat flow $Q$ normalized by $Q_{\text{no slip}}$ (horizontal dashed gray line) evacuated at different flow rate regimes $\dot{V}$ for the different slip-pattern cases $n=25, 50, 100, 200$ and $\theta=45^{\circ}$. The heat flow in the full slip case is shown with gray diamonds.}
  \label{fig:n_variable}
\end{figure}

To quantify the heat flow enhancement observed in Fig.~\ref{fig:angles}, we present in Fig.~\ref{fig:n_variable} the heat flow $Q$ normalized by the standard no-slip case, as a function of $\dot{V}$ and for different numbers of stripes $n$, at a fixed angle $\theta = 45^{\circ}$.
We observe that all the studied patterns show a better performance than the no slip case. Moreover, this improvement is enhanced for larger $\dot{V}$ and increasing $n$, achieving up to $\sim 45\%$ for $n=200$.

\begin{figure}[htbp]
\centering
\begin{subfigure}[b]{0.35\textwidth}
  \centering
  \includegraphics[width=\linewidth]{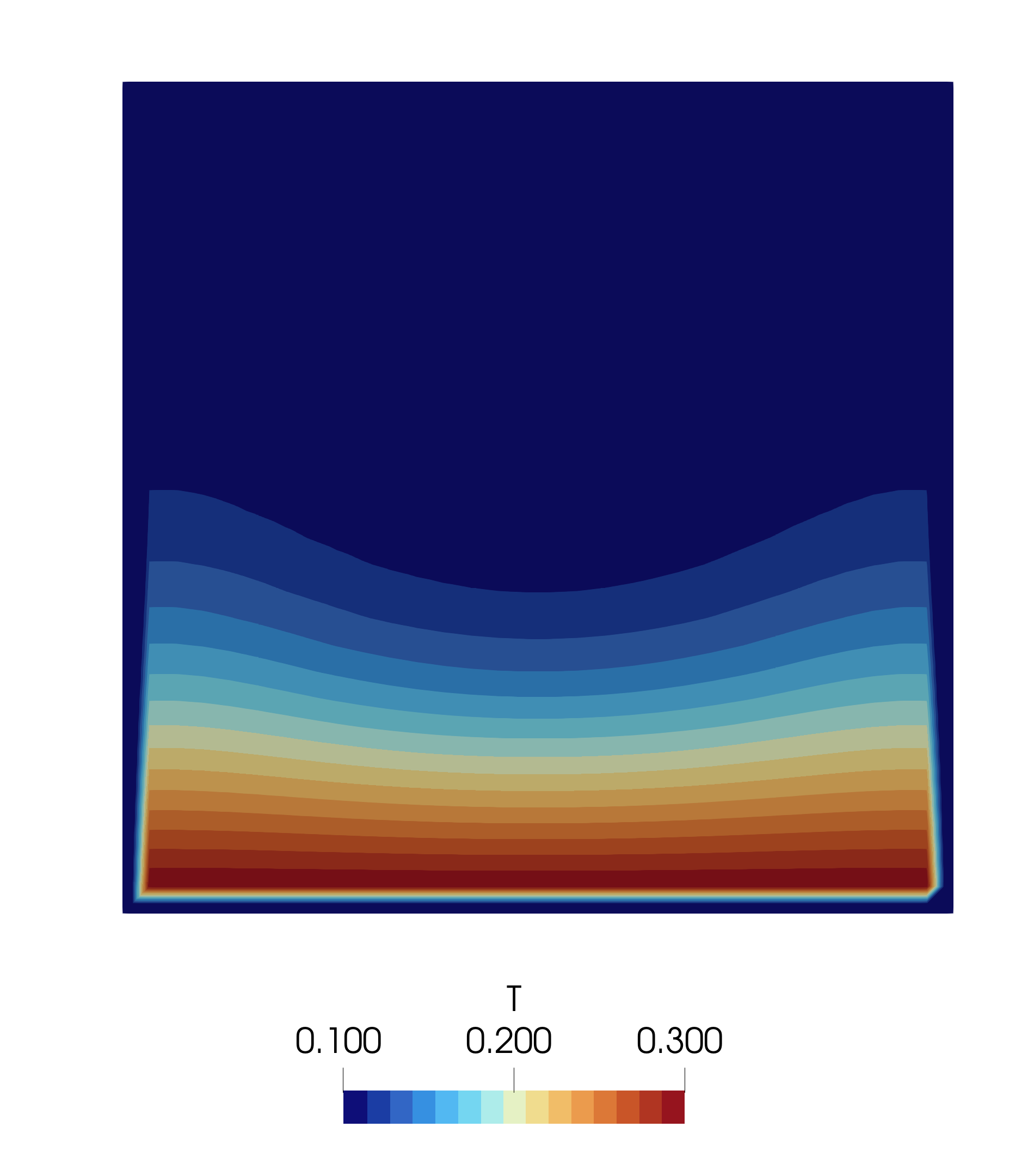}
  \caption{No slip walls}
  \label{fig:no_slip_temp_map}
\end{subfigure}
\begin{subfigure}[b]{0.35\textwidth}
  \centering
  \includegraphics[width=\linewidth]{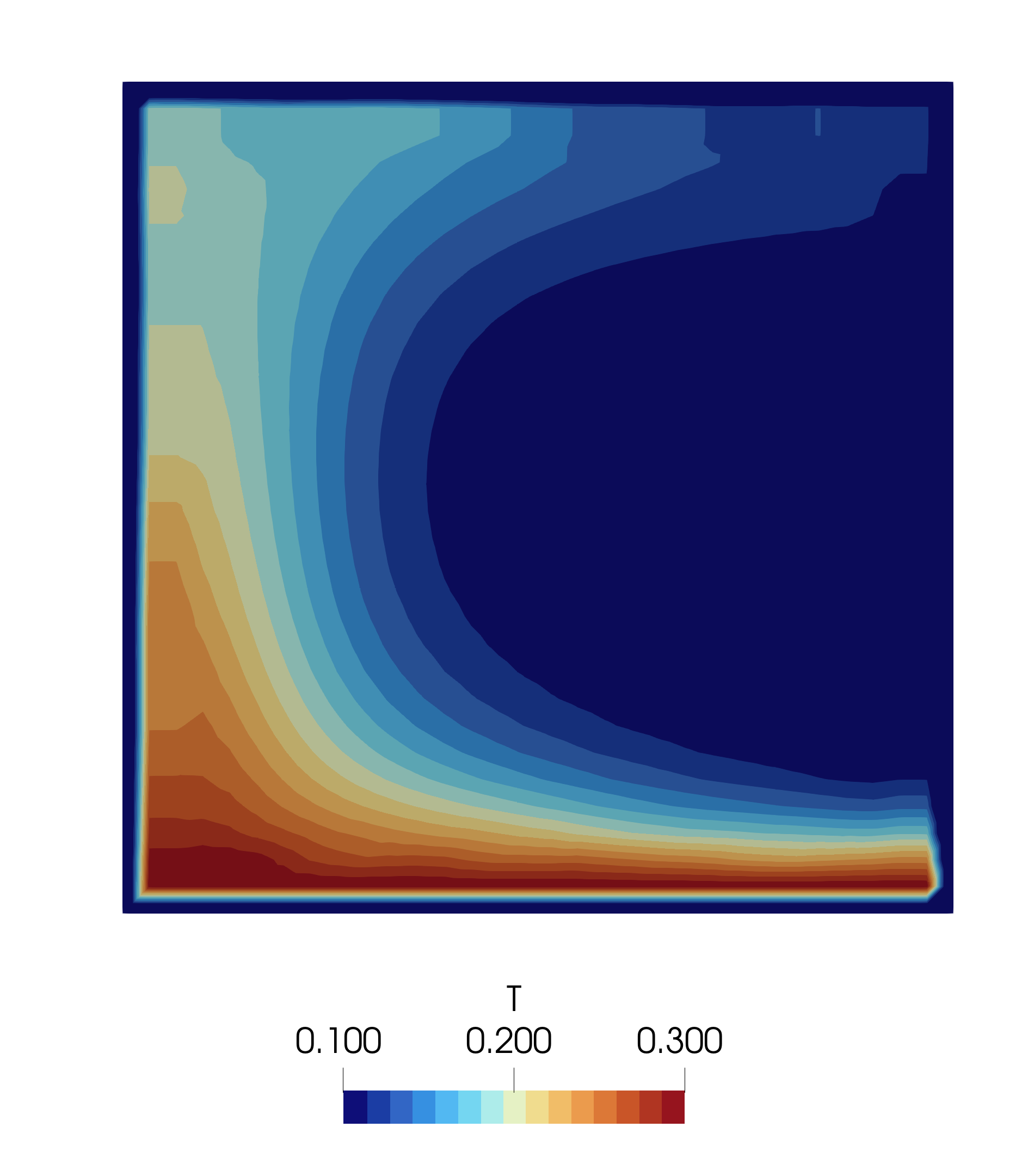}
  \caption{Slip pattern}
  \label{fig:slip_pattern_temp_map}
\end{subfigure}
\caption{Comparison between the temperature cross-sectional profile ($z=L$) for the no slip and the  optimal slip-pattern microchannel ($\theta=45^{\circ}$ $n=200$).}
\label{fig:temp_map}
\end{figure}

To gain a more in-depth understanding of the enhancement of the heat flow we plot in Fig.~\ref{fig:temp_map} the temperature map at the outlet cross-section. We compare the no slip case in Fig.~\ref{fig:no_slip_temp_map} with the case showing the best performance ($n=200$ and $\theta = 45^{\circ}$) in Fig.~\ref{fig:slip_pattern_temp_map}.
We observe that the stripe pattern generates vorticity, that assists heat transfer from the hot bottom wall to the other adiabatic walls.

\begin{figure}[htbp]
    \centering
    \includegraphics[width=\linewidth]{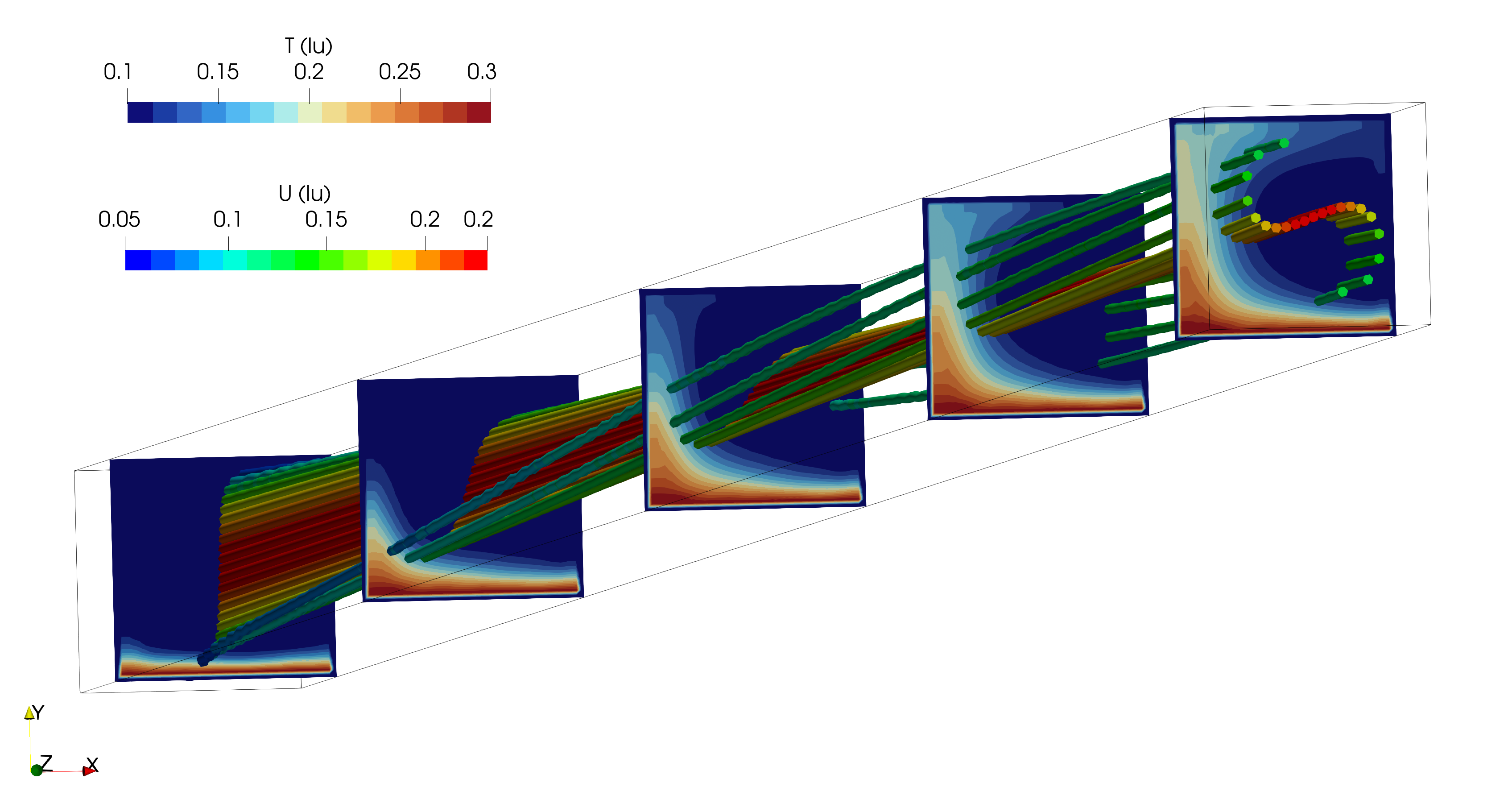}
    \caption{Velocity streamlines and temperature maps at different cross-sectional areas along the channel for the optimal slip pattern case ($\theta=45^{\circ}$ $n=200$).}
    \label{fig:temp_stream_lines}
\end{figure}
To further illustrate this case, in Fig. \ref{fig:temp_stream_lines} we show five cross-sectional slice temperature profiles along the channel superimposed to the velocity stream lines. A helical vortex (swirl) can be observed. 

\begin{figure}[htbp]
    \centering
    \includegraphics[width=0.5\linewidth]{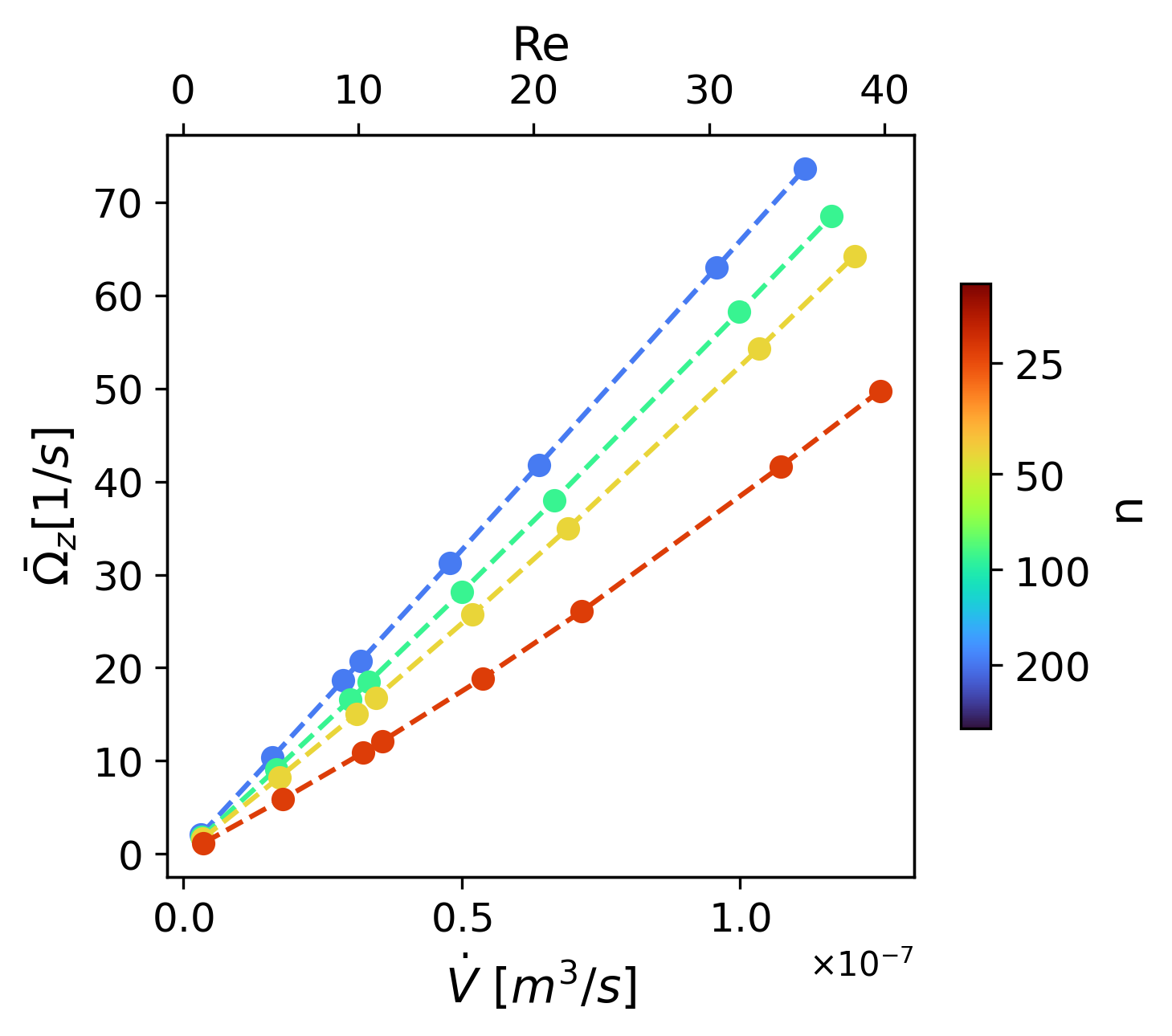}
    \caption{Mean vorticity $\bar{\Omega}_z$ at different flow regimes $\dot{V}$ for the different slip-pattern cases $n=25, 50, 100, 200$ and $\theta=45^{\circ}$}
    \label{fig:vorticity}
\end{figure}

This swirl can be quantified by the mean vorticity $\bar{\Omega}_z$ which is computed as the average over the whole channel volume of the local vorticity given in Eq.~\eqref{eq:vorticity}. Figure ~\ref{fig:vorticity} shows this mean vorticity $\bar{\Omega}_z$ as a function of $\dot{V}$ for different stripe numbers $n$. We observe an almost linear dependence, with increasing slopes of fluid flow ($\dot{V}$) with respect to $n$. Consequently, mean vorticity ($\bar{\Omega}_z$) takes its highest values at an angle of 45° and for $n = 200$ (see Fig. \ref{fig:vorticity}), the configuration that exhibits the best heat transport.

\begin{figure}[htbp]
    \centering
    \includegraphics[width=0.5\linewidth]{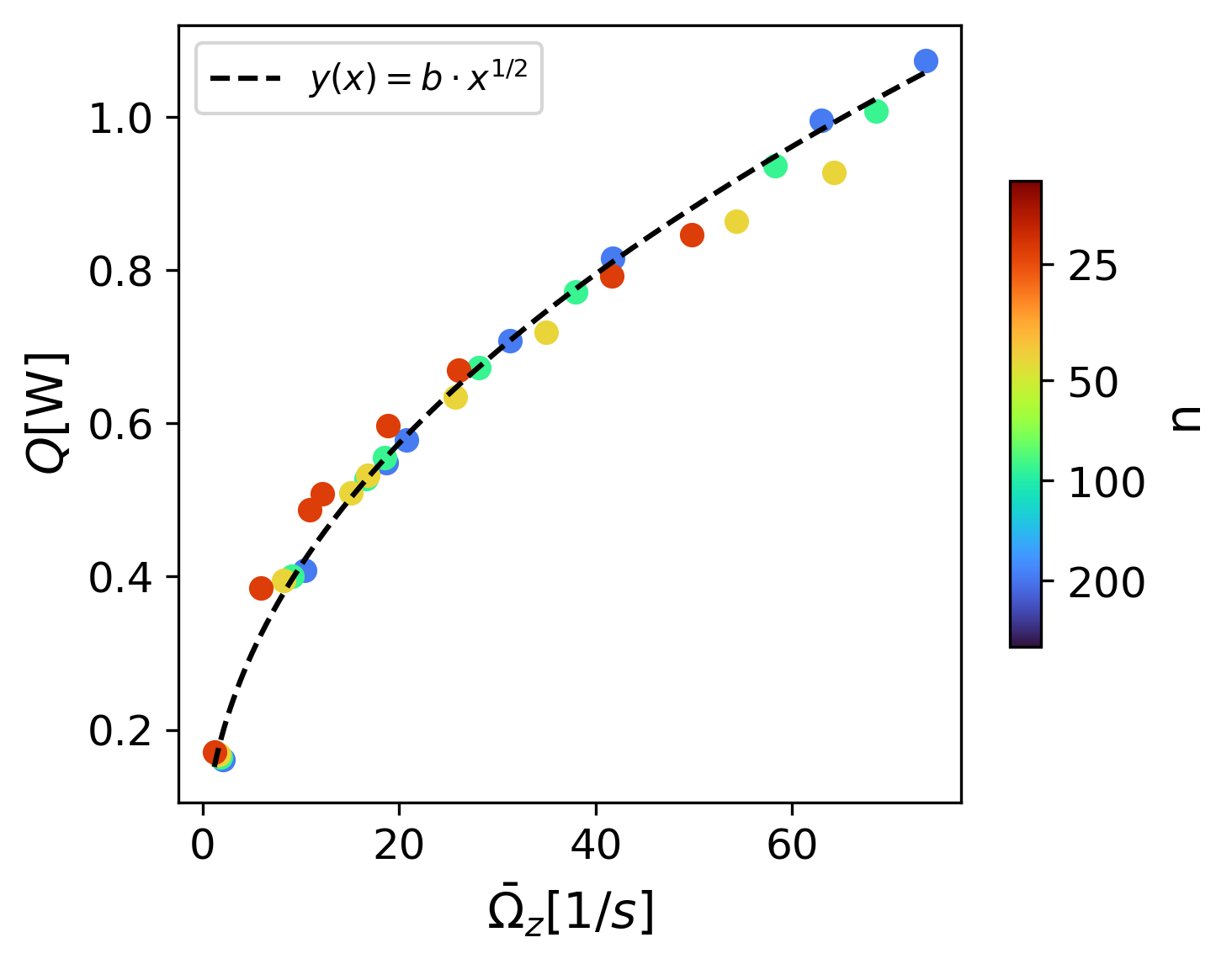}
    \caption{Heat flow $Q$ as a function of vorticity $\bar{\Omega}_z$, for $\theta=45^{\circ}$. Color map indicates the different values of $n$.  Data are fitted with a function $y(x)=b \cdot x^{1/2}$, shown with a dashed black line.  For each $n$ (color), the flow rates increase from left to right as it is shown in Fig. \ref{fig:vorticity}}.
    \label{fig:heatFluxVorticity}
\end{figure}

These results highlight the dependence between the increase in the heat flow and the vorticity induced by the slip pattern, which is a key finding of the study. Fig. \ref{fig:heatFluxVorticity} illustrates this dependence by showing the heat flow ($Q$) as a function of the vorticity ($\bar{\Omega}_z$), for various volumetric flows and patterns. 
Interestingly, all the data points collapse onto a master curve that can be fitted with a power law function compatible with an exponent $1/2$. While we do not have a clear understanding of the functional form of the master curve, the collapse of the data clearly shows that the vorticity number, $\Omega_z$ is the relevant parameter controlling the heat transport along the channel.

\section{Conclusions}\label{sec:conclusions}
In this work we presented a novel strategy to enhance the thermal performance of a MCHS based on slip-no slip pattern designs. Our model considered channels with adiabatic walls, except for the bottom one, which was kept at a constant temperature to mimic a heat source.

We simulated different pattern configurations by varying the number of stripes and their inclination angle. We evaluated the thermal performance of each case at different volumetric flows and compared them to the standard channel under no-slip condition. Our results showed that the underlying mechanism for improving heat transfer is swirling flow produced by the patterning.

Among all the studied pattern configurations, the one with the largest number of stripes and an inclination angle of $45^{\circ}$ showed the largest mean vorticity, with a consequent improvement in the dissipated heat of up to $45\%$ compared to the standard no-slip case.

We also observed a clear correlation between vorticity and heat flux extracted by the microchannel (Fig. \ref{fig:heatFluxVorticity}). This exhibited a power-law dependence with an exponent of $1/2$. However, a complete understanding of this relationship and how the exponent can vary with fluid and geometric parameters, remains an open question for future work.

In situations involving advection-dominated regimes, it is possible to enhance heat removal from localised hot spots by generating swirling flows. These flows can be produced by an appropriate combination of surface patterning, channel geometry and flow rate.
    
Rather than using channels that encompass pillars, posts or ribs, as is often found in the literature, we propose an alternative approach which reduces additional pressure losses, a critical issue for technological implementations. Surface patterning with hydrophobic/hydrophilic properties has emerged as an effective way to improve the thermal performance of microchannel heat sinks.

\bibliographystyle{unsrt}  
\bibliography{swirl}  

\begin{appendices}
\section{Unit Conversion}\label{sec:Appendix I}

Following the literature~\cite{baakeemNovelApproachUnit2021, krugerLatticeBoltzmannMethod2017b}, in this section we detail the conversion units that map quantities between the LB scale Sec. \ref{sec:computational method} and the physical units of the model Sec. \ref{sec:physical model}. In order to do so we need to specify the system dymensions and the properties of the fluid. 

Here we focus on the case of a square channel with cross section of side $w=5\times10^{-4}$m and length $L=50\times w$ with a system temperature range of $15^{\circ}C<T<35^{\circ}C$. As a working fluid we focus on ethylen-glycol/water mixture at $75\%$ volume fraction with constant thermo-physical properties at $25^{\circ}C$~\cite{bohneThermalConductivityDensity1984}: $\rho=1088$ $kg/m^3$, $\mu = 6.85\times 10^{-3}$ $Pa\cdot s$, $\alpha = 1.02\times10^{-7}$ $m^2/s$ and $\kappa = 0.307 $ $W/mK$.

Given the above mentioned physical properties, we choose the conversions of time, length and temperature shown in Table~\ref{table:conv-1}. We note that while for time and length the relations between lattice units and SI units are simply direct proportionality, for the temperature we choose 
\begin{equation}
    T = T_0 + C_T T^{\star}\,,
\end{equation}
being $T_0=278.15$ K a reference temperature and $C_T$ the temperature conversion factor given in Table \ref{table:conv-1} and $0.1<T^{\star}<0.3$.
\begin{table}[h!]
    \caption{Conversion factors for time, length, and temperature}
    \centering
    \begin{tabular}{cccc}\hline\hline \noalign{\smallskip}
        Property            & Conversion Factor                  & Value  \\ 
        \hline
        length              & $C_x = w / w^{\star}$ & $1.56\times 10^{-5}$ m         \\
        time                & $C_t = c_s^{\star 2}
            \left(\tau^{\star}-\frac{1}{2}\right)\frac{C_x^2}{\nu}$ & $3.88\times 10^{-6}$ s         \\
        temperature         & $C_T = \Delta T/ \Delta T^{\star}$ & $100$ K    \\    
        \noalign{\smallskip} \hline\hline 
    \end{tabular}
    \label{table:conv-1}
\end{table}
Using the conversion factors in Table~\ref{table:conv-1}  we derive the conversion factors for other quantities of interest, see Table~\ref{table:conv-2}

\begin{table}[htbp]
    \caption{Conversion factors for other physical magnitudes}
    \centering
    \begin{tabular}{cccc} \hline\hline \noalign{\smallskip}
        Property            & Conversion Factor                  & Value  \\ 
        \hline
        thermal diffusivity & $C_{\alpha}= C_x^2 / C_t$ & $6.29\times 10^{-5}$ m$^2$/s         \\
        aceleration         & $C_a = C_x / C_t^2$ & $1.04 \times 10^{6}$ m/s$^2$         \\
        volume flow rate    & $C_{\dot{V}} = C_x^3 / C_t$ & $9.84\times 10^{-10}$ m$^3$/s         \\
        velocity            & $C_u = C_x / C_t$ & $4.03$ m/s         \\
        convective heat flow            & $C_{Q}=c_p \rho (C_x^3/C_t)C_T$ & $0.29$ W         \\
        \noalign{\smallskip} \hline\hline
    \end{tabular}
    \label{table:conv-2}
\end{table}
Tables~\ref{table:conv-1},\ref{table:conv-2} allow to convert the value of any physical magnitude $Z$ to the corresponding value $Z^*$ in lattice units via $Z^* = Z \,\,  C_Z$, with $C_Z$.

\section{Analytical Solution validation}\label{sec:Appendix II}

We consider a laminar flow in between two infinite planar plates, established by a pressure gradient along the $z$ direction. This situation reduces to an effective 2D problem.

\begin{figure}[htbp]
    \centering
    \includegraphics[width=0.65\linewidth]{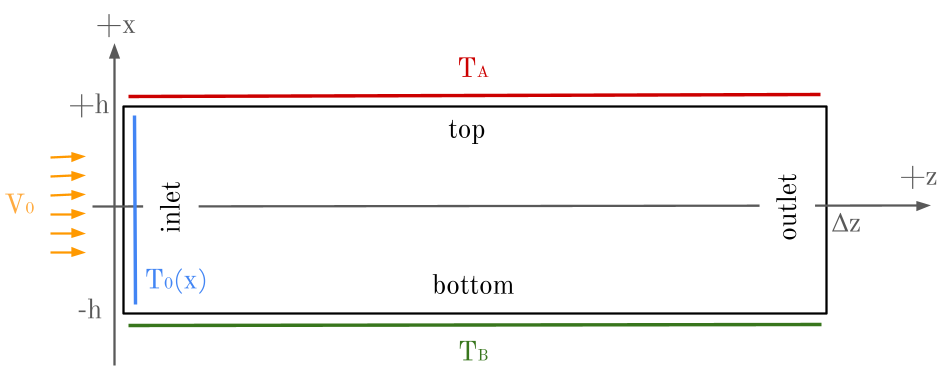}
    \caption{2D physical model diagram }
\label{fig:2d_analytical_layout}
\end{figure}

From the Navier-Stokes equation, imposing velocity in the $z$ direction with a no slip boundary condition in the plates, the
Hagen-Poiseuille solution reads
\begin{equation}\label{eq:hagen-poiseuille}
    u_z (x) = u_{\text{max}}[1-(x/h)^2]\,,
\end{equation}
with
\begin{equation}
 u_{\text{max}} = \frac{1}{2} \frac{\Delta P}{\Delta z} \frac{h^2}{\mu}\,,
\end{equation}

The general advection-diffusion equation for the temperature $T$ into a moving fluid is
\begin{equation}
\rho c_p \left( \frac{\partial T}{\partial t} + \mathbf{u} \cdot \nabla T \right) = \kappa \nabla^2 T + \Phi + Q\,,
\label{advection-diffusion}
\end{equation}
where $\rho$ is the fluid density, $c_p$ is the specific heat capacity at constant pressure, $\mathbf{u}$ is the fluid velocity vector, $\kappa$
is the thermal conductivity of the fluid, 
$\Phi$ is the viscous dissipation term, and $Q$ is a volumetric heat source term.

For a stationary situation, and neglecting heat dissipation and sources, the equation reduces to

\begin{equation}
     \mathbf{u} \cdot \nabla T = \alpha \nabla^2T\,,
\end{equation}
where $\alpha = \kappa/(\rho c_p)$ is the thermal diffusivity of the fluid. For the 2D Poiseuille velocity profile, in Cartesian coordinates we obtain

\begin{equation}
    \left[\frac{\partial^2 T}{\partial x^2} +\frac{\partial^2 T}{\partial z^2} \right] = \frac{u_{\text{max}}}{\alpha}\left[1-\left(\frac{x}{h}\right)^2\right]\frac{\partial T}{\partial z}\,.
    \label{eq:app-B5}
\end{equation}

We have an inlet temperature profile  of the fluid $T_0 (x)$, and constant temperature of each plate $T_{\text{A}}$ and $T_{\text{B}}$, see Fig. \ref{fig:2d_analytical_layout}. Considering these boundary conditions for the temperature we  express the solution as

\begin{equation}
    \label{eq:generalSolution}
    T(x,z) = \sum_{n=1}^{\infty} \gamma_n \ e^{-\lambda_n \frac{z}{h}} \phi_n \left(\frac{x}{h}\right) + T_{\infty} (x)\,,
\end{equation}
with
\begin{equation}
    \label{eq:Tinfinity}
    T_{\infty} (x) = \frac{(T_{\text{A}} - T_{\text{B}})}{2}\frac{x}{h} + \frac{(T_{\text{A}} + T_{\text{B}})}{2}\,.
\end{equation}
The asymptotic solution for $z\rightarrow \infty$, $\phi_n$ and $\lambda_n$ are eigenfunctions and eigenvalues to be determined, and 
$\gamma_n$ are coefficients that will depend on the inlet temperature profile. Replacing into the differential equation:

\begin{align}
    0 = \sum_{n=1}^{\infty} \gamma_n \ e^{-\lambda_n \frac{z}{h}} \ \left[ \frac{1}{h^2} \ \phi_n''\left(\frac{x}{h}\right) + \frac{\lambda_n^2}{h^2} \ \phi_n\left(\frac{x}{h}\right) \right. 
     + \left. \frac{\lambda_n}{h} \frac{u_{\text{max}}}{\alpha} \ \left[1 - \left(\frac{x}{h}\right)^2\right]\ \phi_n\left(\frac{x}{h}\right) \right] \ .
\end{align}
To fulfill this equation, every term in the sum should be zero, arriving to a differential equation

\begin{equation}
\label{ParabolicDiff}
    0 = \phi''(\eta) + \lambda^2 \phi(\eta) + \lambda \beta \ ( 1 - \eta^2) \ \phi (\eta)\,,
\end{equation}
with $\eta = x/h$ the dimensionless transversal $x$ coordinate, and $\beta = h u_{\text{max}} / \alpha$ a dimensionless parameter proportional to the maximum velocity. For the proposed solution Eq. (\ref{eq:generalSolution}) to give a constant temperature at $x = \pm h$, the functions $\phi$ should fulfill the boundary conditions:

\begin{equation}
    \phi (1) = \phi (-1) = 0 \,. 
\end{equation}
Eq. (\ref{ParabolicDiff}) is a parabolic differential equation with two independent solutions
\begin{align}
    \psi_1 (\beta, \lambda, \eta) &= e^{-\frac{1}{2} \sqrt{\beta \lambda} \, \eta^2} \, F_1 \left( \frac{1}{4} - a(\beta, \lambda); \frac{1}{2}; \sqrt{\beta \lambda} \, \eta^2 \right)\,, \\
    \psi_2 (\beta, \lambda, \eta) &= \eta \, e^{-\frac{1}{2} \sqrt{\beta \lambda} \, \eta^2} \, F_1 \left( \frac{3}{4} - a(\beta, \lambda); \frac{3}{2}; \sqrt{\beta \lambda} \, \eta^2 \right)\,,
\end{align}
where ${}_1F_1$ is the Kummer confluent hypergeometric function\cite{abramowitzHandbookMathematicalFunctions1972}, with
\begin{equation}
    a (\beta, \lambda) = \frac{1}{4} (\beta + \lambda) \sqrt{\frac{ \lambda }{\beta }}\,.
\end{equation}
These two solutions depend parametrically on $\beta$ and $\lambda$, being even and odd functions of $\eta$, respectively.
The general solution to Eq. (\ref{ParabolicDiff}) is

\begin{equation}
    \phi (\eta) = D_1 \psi_1 (\beta, \lambda, \eta) + D_2 \psi_2 (\beta, \lambda, \eta)\,.
\end{equation}
In order to fulfill the temperature boundary conditions (see right after Eq.~\eqref{eq:app-B5}) we need
\begin{align}
    \phi(1)  &= 0 = D_1 \, \psi_1 (\beta, \lambda, 1) + D_2 \, \psi_2 (\beta, \lambda, 1)\,, \\
    \phi( -1) &= 0 = D_1 \, \psi_1 (\beta, \lambda, -1) + D_2 \, \psi_2 (\beta, \lambda, -1)\,.
\end{align}
For non trivial solutions, the determinant of this linear homogeneous system 
\begin{align}
    \det (\beta, \lambda) = \psi_1 (\beta, \lambda, 1) \ \psi_2 (\beta, \lambda, -1)\, 
     - \psi_1 (\beta, \lambda, -1) \ \psi_2 (\beta, \lambda, 1)\,,
\end{align}
must be zero.
Fixing a specific value for $\beta$, this determinant is an oscillating function of $\lambda$ with infinitely many zeros. We numerically compute the first ones, ordering them by increasing values $\{ \lambda_1, \lambda_2, \ldots, \lambda_n \}$, being the eigenvalues with the corresponding eigenfunctions
\begin{equation}
    \phi_n (\eta) = \psi_2 (\beta, \lambda_n, 1) \ \psi_1( \beta, \lambda_n, \eta)  - \psi_1(\beta, \lambda_n, 1) \ \psi_2 (\beta, \lambda_n, \eta)\,.
\end{equation}
We consider that the given inlet temperature profile $T_0 (x) = T (x, z = 0)$ fulfills the conditions $T_0 (h) = T_{\text{A}}$ and $T_0 (-h) = T_{\text{B}}$, to avoid discontinuities of the temperature at the walls.

Numerical inspections revealed that the eigenfunctions $\phi_n$ are not orthogonal on the domain $\eta = [-1,1]$.

\begin{figure}[htbp]
    \centering
    \includegraphics[width=0.5\linewidth]{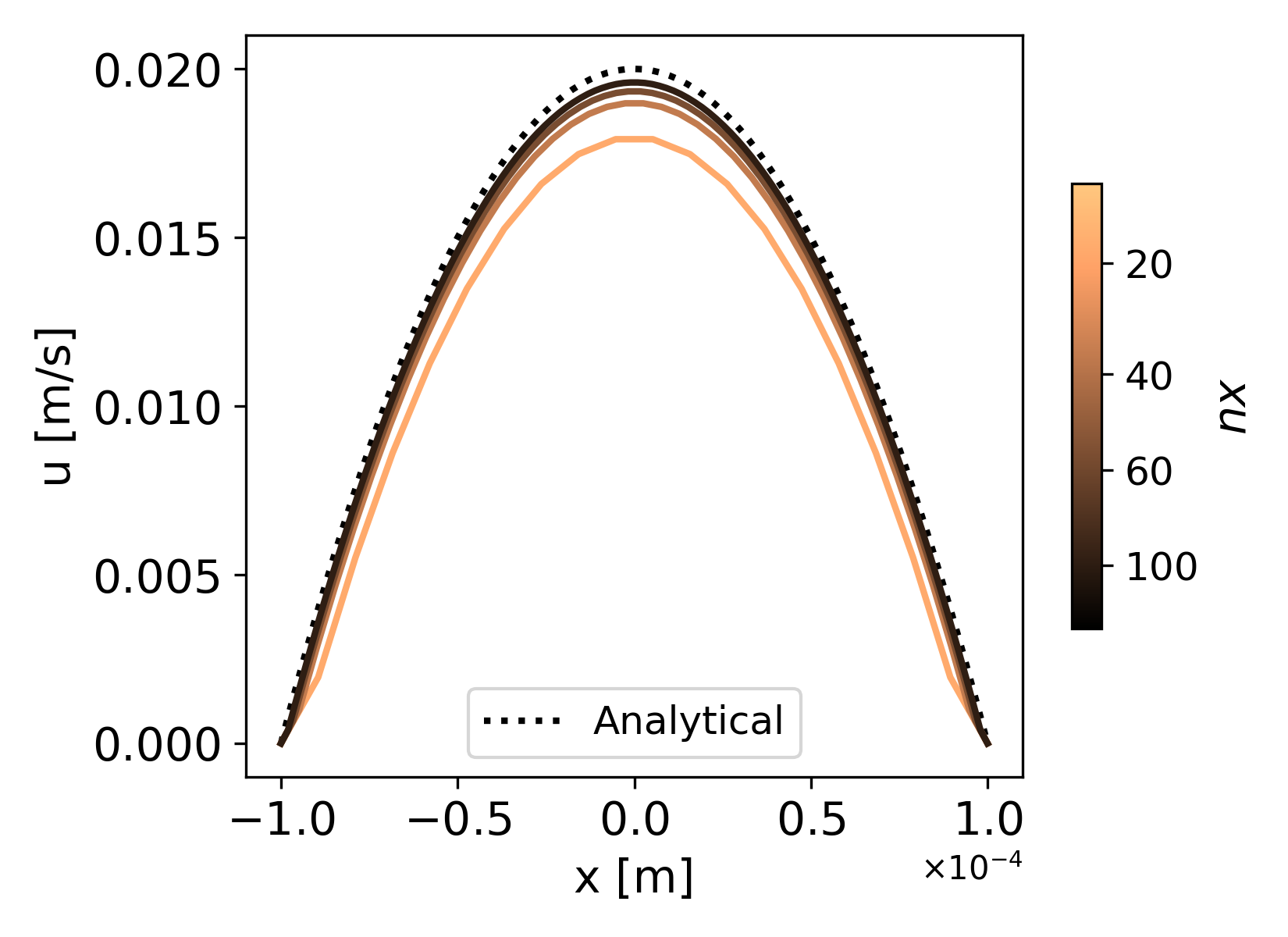}
    \caption{Velocity profile for different mesh sizes ($nx=$20, 40, 60 and 100) compared to the analytical Hagen-Poiseuille  Eq.~\ref{eq:hagen-poiseuille} (dotted line).}
    \label{fig:vel_check}
\end{figure}
\begin{figure}[htbp]
    \centering
        \begin{subfigure}[b]{0.45\textwidth}
          \centering
          \includegraphics[width=\linewidth]{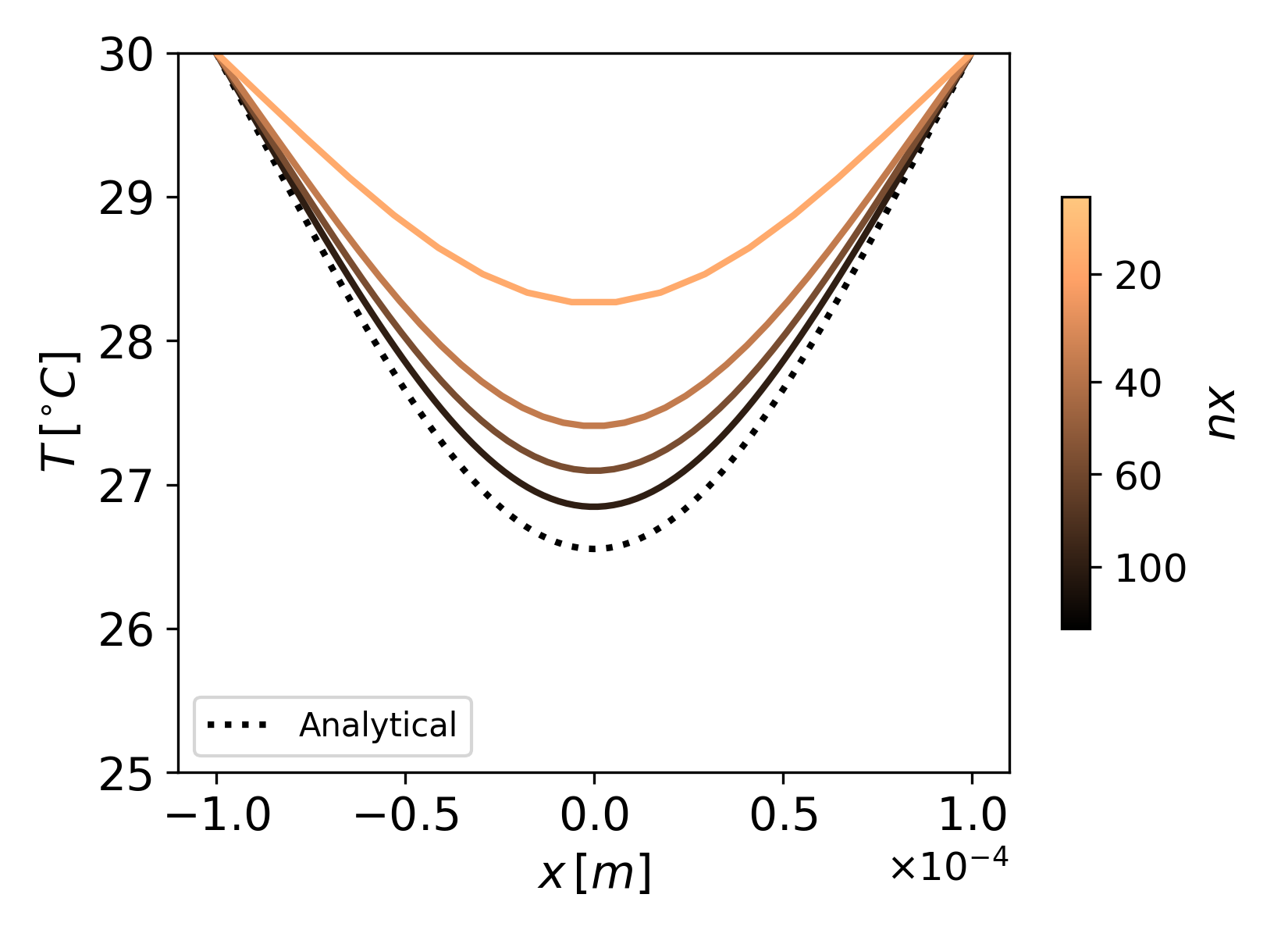}
          \caption{Temperature profile along channel height at $z=\Delta z/2$}
          \label{fig:temp_analytical_x}
        \end{subfigure}
        \begin{subfigure}[b]{0.45\textwidth}
          \centering
          \includegraphics[width=\linewidth]{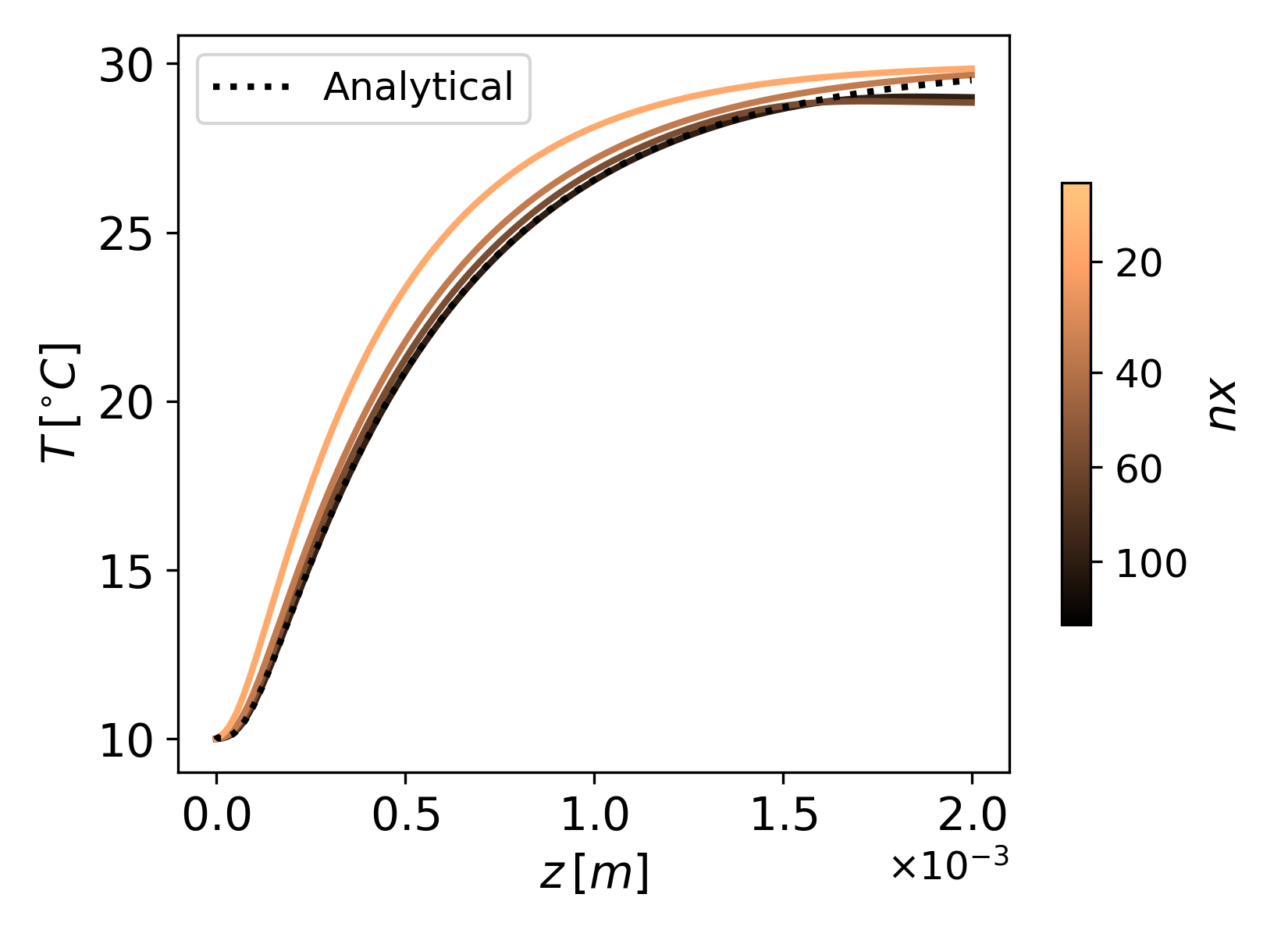}
          \caption{Temperature profile along channel length at $x=0$}
          \label{fig:temp_analytical_z}
        \end{subfigure}
    \caption{Temperature profiles for different mesh sizes ($nx=$20, 40, 60 and 100) compared to the analytical solution Eq.~\eqref{eq:generalSolution} (dotted line).}
    \label{fig:temp_check}
\end{figure}
Therefore, to obtain the coefficients $\gamma_n$ we have to truncate the basis up to some maximum value $N$, solving the linear system

\begin{equation}
    \boldsymbol{A} \cdot  \boldsymbol{\gamma} =  \boldsymbol{B}\,,
\end{equation}
with the matrix $\boldsymbol{A}$ with elements

\begin{equation}
    A_{i j} = \int_{-1}^1 \phi_i (\eta) \ \phi_j (\eta) \ \text{d} \eta\,,
\end{equation}
and the vector
\begin{equation}
    B_{j} = \int_{-1}^1 f_0 ( \eta) \ \phi_j (\eta) \ \text{d} \eta\,,
\end{equation}
with $f_0 ( \eta) = T_0 (h \eta) - T_{\infty} (h \eta)$, a function that it is zero for $\eta= -1$ and $\eta= 1$.
With the computed eigenvalues $\lambda_n$, eigenfunctions $\phi_n$ and the $\gamma_n$ coefficients, we can replace them into the general solution (\ref{eq:generalSolution}), and we obtain the temperature map of the fluid for $z > 0$. 

We use these analytical results to validate the thermal calculations with the LBM. For this purpose, the following physical quantities are taken into consideration:
pressure difference $\Delta P = 8$ Pa, $H=2h=2\times10^{-4}$ m, and $\Delta z = 0.002$ m. The acceleration $a = \Delta P / (\Delta z\,\rho)=4$m/s$^2$ . We consider thermophysical properties of pure water at $20^{\circ}$C: density $\rho = 1000$ kg/m$^3$, dynamic viscosity $\mu = 10^{-3}$ Pa.s, and thermal diffusivity $\alpha = 1.4\times 10^{-7}$m$^2$/s.
The temperature boundary conditions at the plates are $T_{\text{A}} = T_{\text{B}} = 30^{\circ}$ C, and the inlet temperature is 
\begin{equation*}
    T_0(x) = 10^{\circ} \text{C} + 20^{\circ} \text{C} \left( \frac{x}{h} \right)^{12}\,,
\end{equation*}
With these values the maximum velocity $u_{max}= 0.02$ m/s, which gives $\beta = 14.28$ for the analytical solution.

Regarding the Lattice Boltzmann quantities, a mesh convergence study was performed by testing various grid resolutions with  $nx = 20, 40, 60,$ and $100$, the number of lattice nodes in x-axis direction. The lattice velocity of sound is  $c_s^* = \sqrt{1/3}$, and the relaxation time is $\tau^* = 1$. The total height of the channel in lattice units is $H^{\star} = nx$ ($h^{\star}=nx/2$). The density is fixed at $\rho^* = 1$.

The procedure was carried out as follows:
\begin{enumerate}
    \item The transverse grid size $nx$ is defined (20, 40, 60, 100), and the longitudinal size is set to $nz = 10\,nx$. The spatial resolution is then computed as $C_x = H/nx$.
    \item The relaxation time $\tau^{\star}=1$, the lattice speed of sound $c_s^{\star}=\sqrt{1/3}$, and given the kinematic viscosity $\nu$, the time conversion factor $C_t$ is obtained according to Eq. \ref{eq:time_conv_factor}.
    \item The force density $F^{\star} = a^{\star}\rho^{\star}$ used to run the simulation is calculated using $C_x$, $C_t$ and $a$.
    \item The lattice temperature boundary conditions are $T^{\star}_A=T^{\star}_B=0.3$ and $T^{\star}_0=0.1$, obtained using the conversion factors from Table \ref{table:conv-1}.  Additionally, the lattice thermal diffusivity $\alpha^{\star}$ is calculated (See Table \ref{table:conv-2}).
\end{enumerate}

Finally, the numerical results for each $nx$ are converted to MKS units, as detailed in Appendix \ref{sec:Appendix I}. These results are compared with the analytical solutions presented in Eqs. \ref{eq:hagen-poiseuille} and \ref{eq:generalSolution}. Figures \ref{fig:vel_check} and \ref{fig:temp_check} show a good agreement achieved between the numerical results and the analytical model as the lattice resolution increases.

\section{Stripe angle variation study}\label{sec:Appendix III}

We present Q vs $\dot{V}$ for the different pattern configurations corresponding to $n = 25, 100, 150, 200$, as well as three stripe angles $\theta = 25^{\circ}, 45^{\circ}, 65^{\circ}$.
For reference, the homogeneous no-slip case (open
diamonds) and the homogeneous full slip case (filled diamonds) are also plotted in each Figure. As can be seen, in each of the four cases, $\theta =45^{\circ}$ corresponds to the best configuration.

\begin{figure*}[h]
    \centering
    \includegraphics[width=0.45\textwidth]{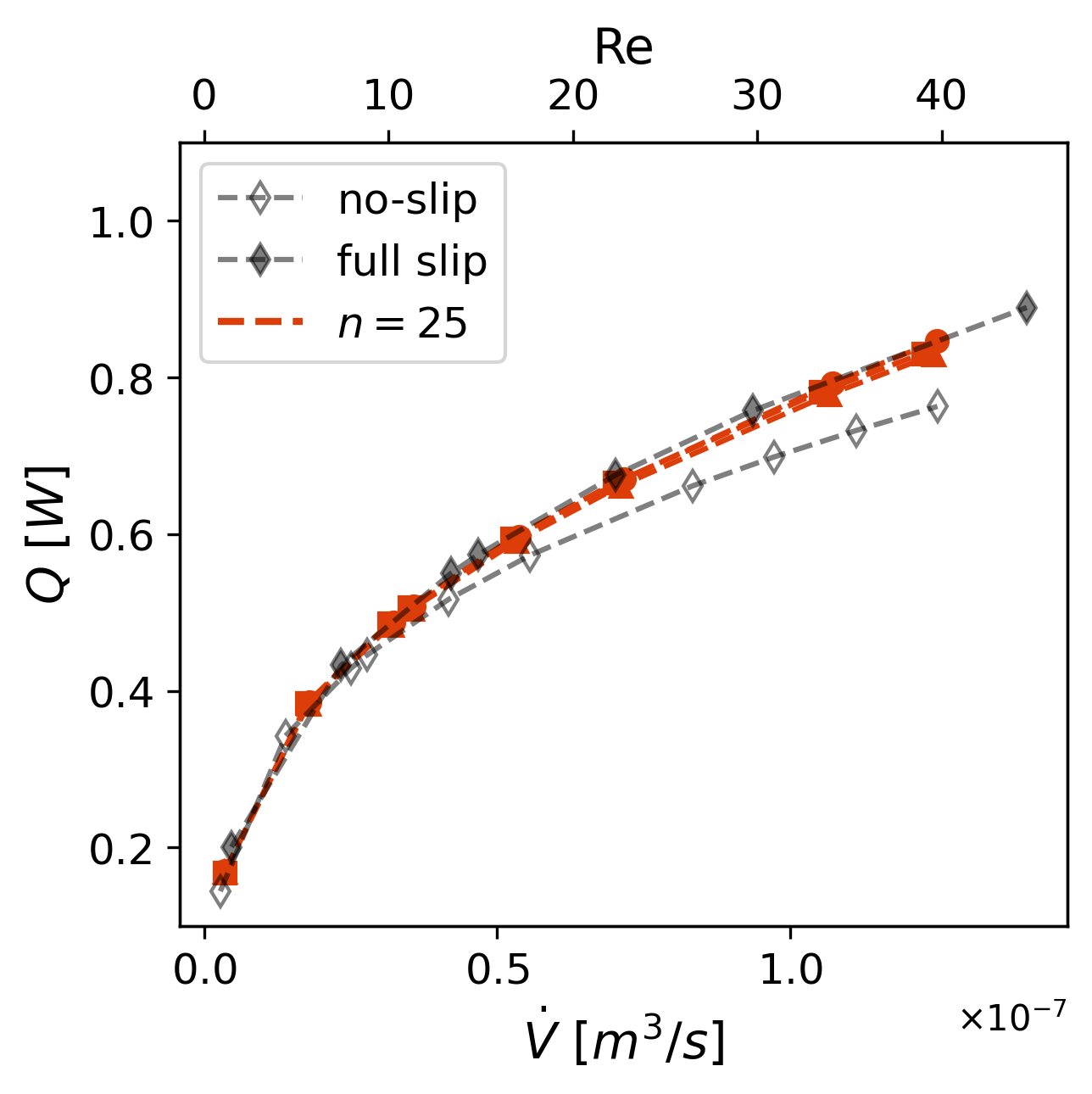}
    \includegraphics[width=0.45\textwidth]{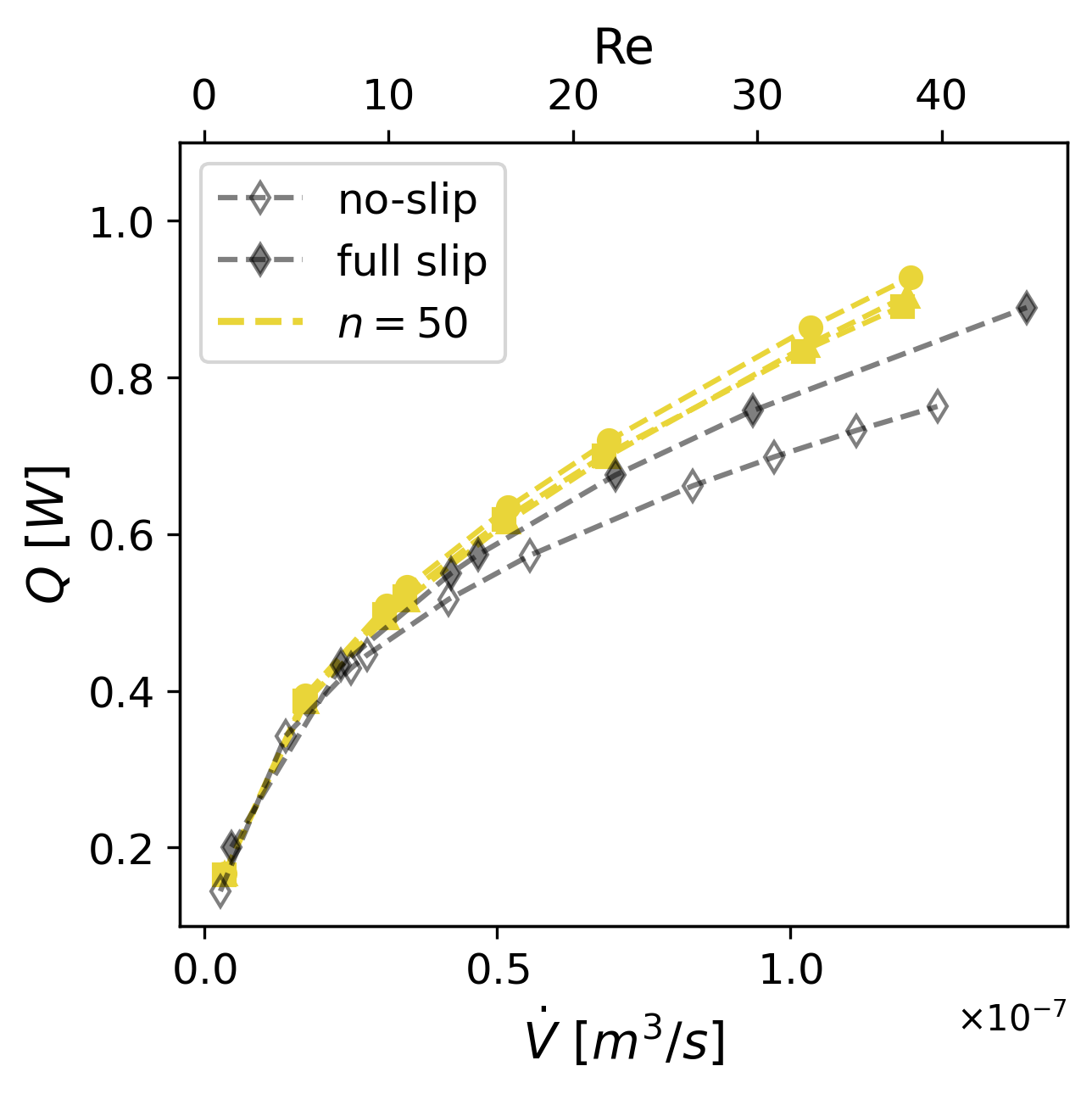}
    \includegraphics[width=0.45\textwidth]{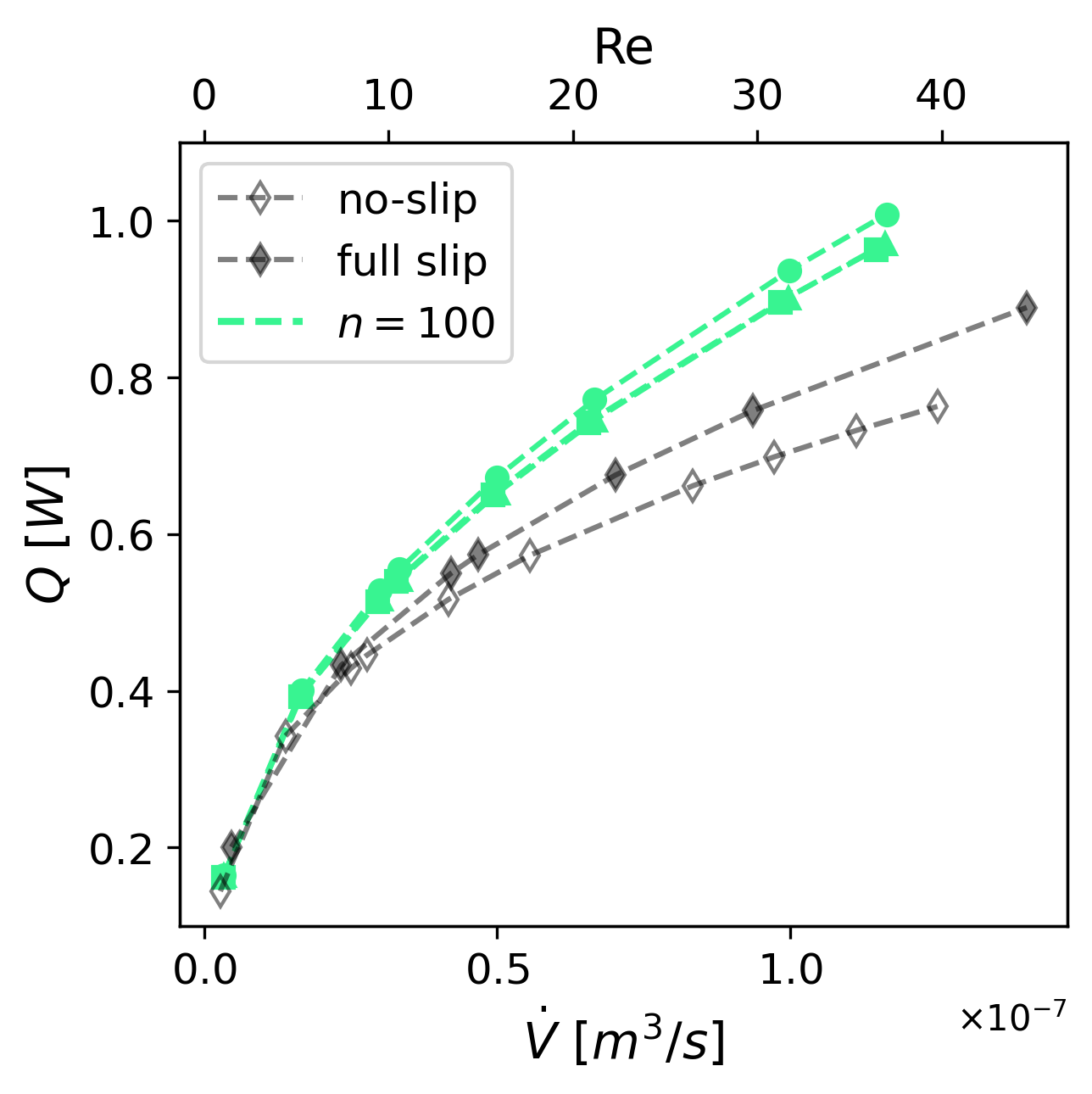}
    \includegraphics[width=0.45\textwidth]{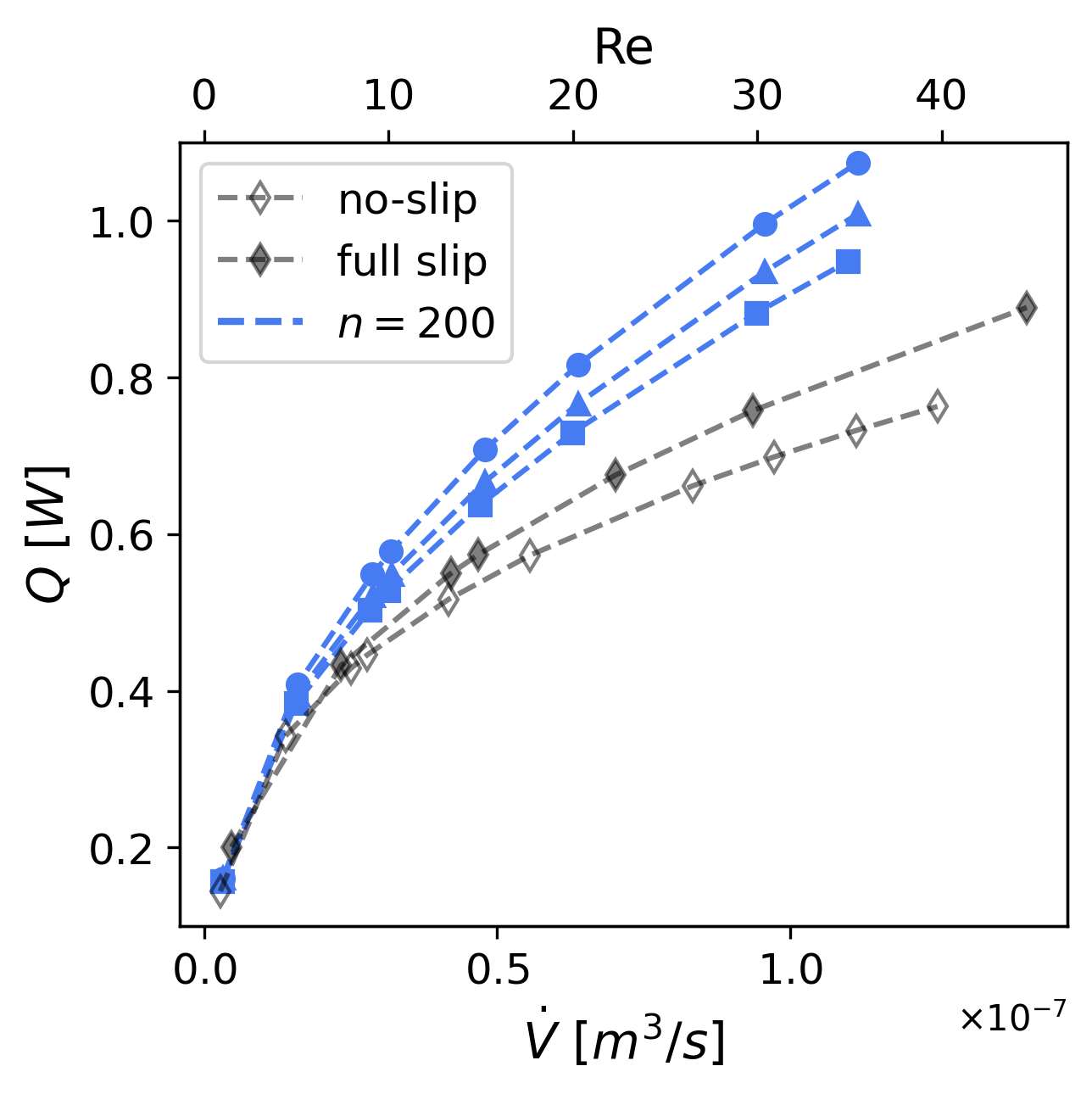}
    \caption{Heat flow evacuated at different flow rate regimes varying the stripe angle $\theta$ at the channel walls (triangles for $\theta=25^{\circ}$, circles for $\theta=45^{\circ}$, and squares for $\theta=65^{\circ}$) for each stripe number case ($n=25$, $50$, $100$ and $200$).}
    \label{fig:app-C}
\end{figure*}

\end{appendices}

\end{document}